\documentclass[fleqn,10pt]{wlscirep}
\usepackage[utf8]{inputenc}
\usepackage[T1]{fontenc}
\title{Optical vortices by an adaptive spiral phase plate}

\author[1]{T. Jankowski}
\author[2]{N . Bennis}
\author[2] {P. Morawiak}
\author[3]{D. C. Zografopoulos}
\author[2,4]{A. Pakuła}
\author[5]{M. Filipiak}
\author[5]{M. Słowikowski}
\author[6,7,8]{J. M L{\'o}pez-Higuera}
\author[6,7,8,*]{J. F. Algorri}

\affil[1]{Faculty of Mechatronics, Warsaw University of Technology, {\'S}w. Andrzeja Boboli 8, 02-525 Warsaw, Poland}
\affil[2]{Faculty of Advanced Technologies and Chemistry, Military University of Technology, Warsaw 00-908, Poland}
\affil[3]{Consiglio Nazionale delle Ricerche, Istituto per la Microelettronica e Microsistemi (CNR-IMM), Roma 00133, Italy}
\affil[4]{Faculty of Mechatronics, Warsaw University of Technology, {\'S}w. Andrzeja Boboli 8, 02-525 Warsaw, Poland}
\affil[5]{The Centre for Advanced Materials and Technologies CEZAMAT, Warsaw University of Technology, 19 Poleczki St., 02-822 Warsaw, Poland}
\affil[6]{Photonics Engineering Group, University of Cantabria, 39005, Santander, Spain}
\affil[7]{CIBER de Bioingeniería, Biomateriales y Nanomedicina, Instituto de Salud Carlos III, 28029, Madrid, Spain}
\affil[8]{Instituto de Investigaci{\'o}n Sanitaria Valdecilla (IDIVAL), 39011, Santander, Spain}

\affil[*]{algorrijf@unican.es}


\begin{abstract}
An Adaptive Spiral Phase Plate (ASPP) based on liquid crystal (LC) and the transmission electrode technique is theoretically and experimentally demonstrated. This ASPP design enables the generation of high-quality optical vortices with topological charges ranging from $\pm1$ to $\pm4$ using a single device (but using a higher birefringence LC and thickness this number can be multiplied by four). The continuous reconfigurability of the optical phase shift, achieved through a simple control mechanism involving only two low voltages, sets this device apart as the most accurate approximation to an ideal ASPP proposed to date. This device offers remarkable advantages, such as complete reconfigurability, allowing adjustment of operating wavelengths and topological charges. The fabrication process mirrors that of a standard LCD cell, ensuring a cost-effective and reliable solution. Its versatile applications, including fiber optics communications and atom manipulation, promise reduced fabrication costs for existing devices and the generation of diverse Orbital Angular Momentum (OAM) modes. In summary, the proposed ASPP stands as a pivotal advancement, providing superior light efficiency, simplicity, and the capability for on-the-fly reconfiguration in a variety of optical applications.

\end{abstract}
\begin{document}

\flushbottom
\maketitle

\thispagestyle{empty}

\section*{Introduction}
Over the last decade, helical wavefronts have been amongst the most extensively studied complex phase-shapes of light. They are characterized by an azimuthal phase dependence of $e^{-il\Phi}$, where $l$ is a variable that represents the topological charge of the vortex, which determines how many times the phase of the wavefront wraps around in 360\textdegree, and $\Phi$ is the azimuthal angle in polar coordinates, which measures the angle in the plane perpendicular to the direction of propagation of the light beam. When focused, these light beams carry an orbital angular momentum (OAM) that forms rings rather than points in a focal plane. 

The potential and applications of this phenomenon have seen an exponential rise with use-cases ranging from laser processing \cite{hamazaki2010, toyoda2012,omatsu2019}, beam shaping \cite{padgett2011}, optical tweezers \cite{ladavac2004}, atom manipulation \cite{tabosa1999} and free-space communications \cite{wang2012}. The last decade also witnessed rapid advancements in the applications of optical vortices and OAM. In 2013, the potential for terabit-level high-capacity optical communication was demonstrated via OAM multiplexing in fibres \cite{bozinovic2013}. Moving onto 2016, in Ref. \cite{fickler2016} demonstrated the generation of extreme OAM states exceeding 10.000$\hbar$ and accomplished quantum entanglement of these states. In the following three years, a plethora of tunable properties of optical vortices have been skillfully managed at the nanoscale. These include the conversion of spin angular momentum to OAM for classical  \cite{devlin2017} and quantum light  \cite{stav2018}, tunable wavelengths ranging from visible light \cite{kong2017,gauthier2017} to X-ray light  \cite{lee2019}, ultra-broadband tunable OAM \cite{xie2018}, adaptable chirality \cite{zambon2019} and time-varying OAM in extreme-ultraviolet vortex beams \cite{rego2019}. Other remarkable applications have been high optical power lenses \cite{geday2020} or advanced microprinting technology \cite{yuyama2023}. 

Up to the present time, optical vortices have sparked countless innovative applications in numerous fields, with their capacities for novel applications continually expanding due to advancements in tunability. Traditionally, optical vortices are generated using spiral phase plates (SPPs) \cite{beijersbergen1994} or computer-generated holograms (CGHs) \cite{heckenberg1992}, with the latter created digitally by computing holographic interference patterns. Blazed forked gratings have been instrumental in concentrating more energy in the first diffraction order \cite{zeylikovich2007}. However, fabricating CGHs remains a challenge. The use of SPPs also faces drawbacks due to fabrication limitations and a lack of reconfigurability, leading to fixed topological charges and operating wavelengths. Recent years have seen attempts to resolve these issues through alternatives like spatial light modulators (SLMs) \cite{crabtree2004} (with an efficiency around $60\%-85\%$ range \cite{forbes2016} and poor fill factors $<40\%$\cite{sanchez2023}) and novel devices using liquid crystal (LC) and individual pie slices of indium tin oxide (ITO) as contact electrodes \cite{albero2012,canogarcia2018,pereiro-garcia2023} (a comprehensive review can be found in Ref. \cite{sanchez2023}). Yet, these solutions present their own set of challenges, such as complex voltage control, light efficiency loss and complications due to the need to combine the phase pattern with a blazed grating pattern.

In this context, the present study introduces a breakthrough structure employing LC and the transmission electrode technique. Previous research has showcased the application of this method in spatial phase modulation of several types \cite{algorri2020, zemska2023} and in various kinds of LC lenses, e.g., cylindrical and Powell \cite{algorri2020b,feng2022}, axicons\cite{algorri2020c,algorri2020c,stevens2022}, aspherical \cite{bennis2022,algorri2020d,pusenkova2022,feng2023,feng2023b} and arrays \cite{feng2022b}. In this scenario, by leveraging a variation of the transmission electrode technique, it is feasible to attain a spiral voltage profile. This innovative structure can generate a continuous spiral phase profile that is completely reconfigurable using only two low voltages. This novel adaptive spiral phase plate (ASPP) has the potential to generate various positive and negative topological charges (up to $l = \pm 4$ are here experimentally demonstrated with a low birefringence LC), reducing complexity and enhancing efficiency in applications, such as optical tweezers or OAM mode division multiplexing, paving the way for the next wave of advancements in the exciting field of optical vortices.

\section*{Structure and operation principle}
	
\begin{figure}[ht]
\centering\includegraphics[width=14cm]{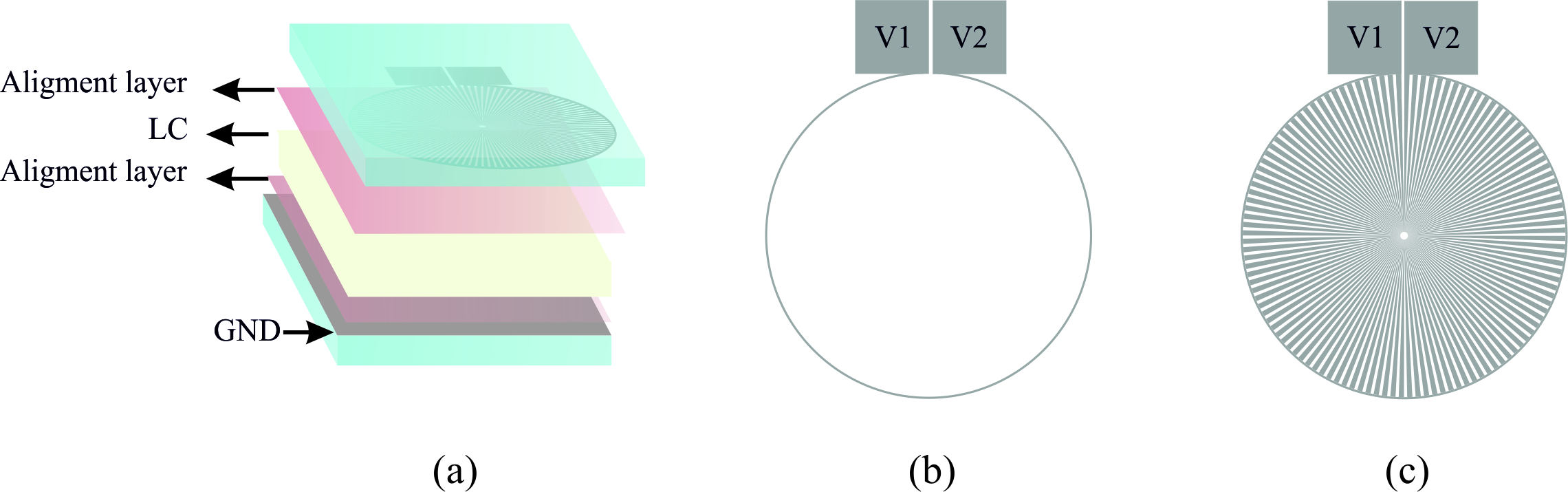}
\caption{(a) Schematic depiction of the LC-tunable optical vortex generator and its various constituent parts. (b) Schematic depiction of the circular transmission electrode alone and (c) including the ITO slices evenly arranged to distribute the voltage. Note these drawings are not to scale (the space between adjacent slices is only 10~{\textmu}m).}\label{vortexelectrode}  
\end{figure}

The device comprises a pair of substrates, each coated with ITO. Only one substrate is photolithographed with the transmission electrode pattern. Fig.~\ref{vortexelectrode}(a) illustrates a schematic of the overall structure. The substrates are separated by a $50$-{\textmu}m thick spacer. The cavity formed in between is occupied by the nematic LC 6-CHBT, with a birefringence $\Delta n = 0.16$, dielectric permittivity $\Delta\epsilon = 7$ \cite{dabrowski1984} and elastic constants $K_{11} = 6.71$~pN (splay), $K_{22} = 2.93$~pN (twist), and $K_{33} = 6.1$~pN (bend) \cite{buchecker1987}. The device is designed as an adaptive phase-only modulator and does not require molecular twist. Hence, homogeneous alignment is made by a rubbing process. This is achieved by utilising a polyimide alignment layer (Nissan SE-130), rubbed parallel to the electrodes for the upper glass and in the opposite direction for the lower one.

The operational principle of the device is straightforward. The key component is a circular transmission electrode (radius $r = 1$~cm) that generates a continuous voltage profile from one electric contact to the other, the grey square pads of Fig.~\ref{vortexelectrode}(b).  In this case, an electrode width ($w$) of $10$~{\textmu}m is chosen for the transmission electrode. The voltage values $V_1$ and $V_2$ represent the applied voltage on the top electrodes. Subsequently, a series of ITO pie slices ($N= 100$ slices in this case) that extend from the periphery to the centre distributes the voltage throughout the active area, see Fig.~\ref{vortexelectrode}(c). Note that drawings are not to scale, the actual gap between slices is $g = 10$~{\textmu}m. The broader part of the slices has $2\pi r/100 - g = 618$~{\textmu}m, whereas it ends on a spike. One of the advantages of the transmission electrode technique is that the resistance from $V_1$ to $V_2$ is very high. Considering the sheet resistance ($R_\textrm{sq} = 100 \Omega$/sq) and ITO thickness ($t=20$~nm), the theoretical resistance (based on numerical simulations) is $R=66.4$~k$\Omega$. For example, for gradient voltages of $1$~V$_\textrm{RMS}$ it implies a current of only $15$~{\textmu}A$_\textrm{RMS}$. On the other hand, one challenge of this configuration is that a certain central area is unused due to the gap between slices. Specifically, a circumference with a radius $r = N g/2\pi = 159$~{\textmu}m. As observed in the simulations and experimental results, this region is very small compared to the active area. Moreover, it could be considerably reduced by improving the resolution of the photolithography. For example, a $1$~{\textmu}m resolution photolithography would produce only a $15.9$~{\textmu}m singularity region that, being considerably lower than the thickness, would lead to a low voltage/phase step.

\begin{figure}[ht]
\centering\includegraphics[width=12cm]{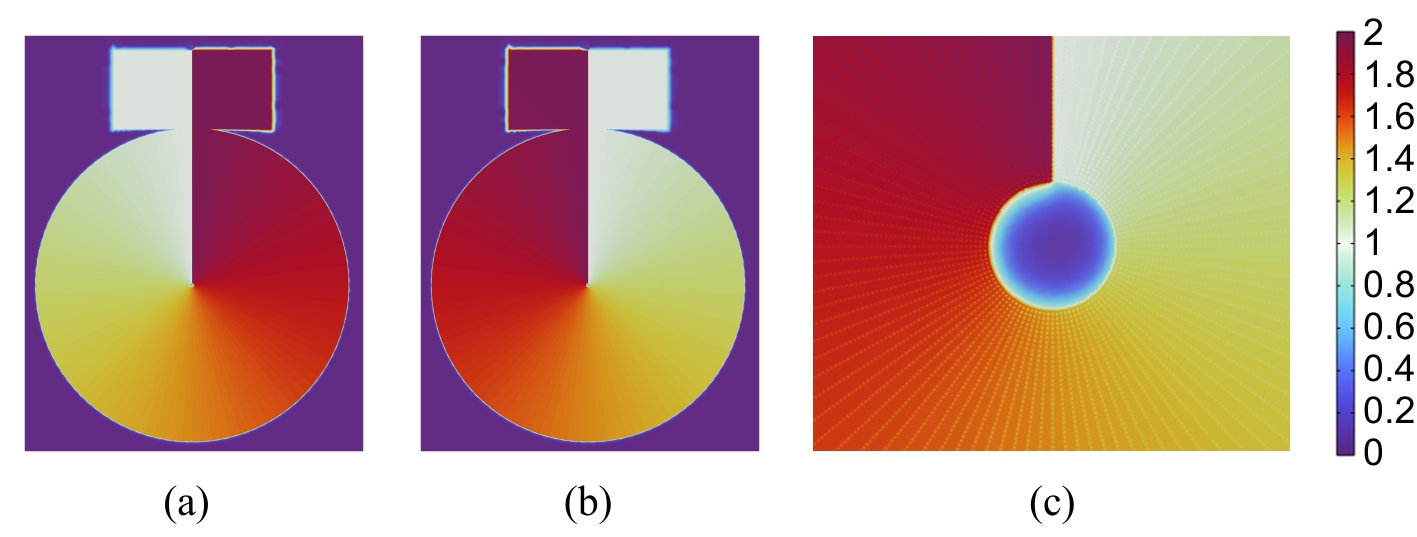}
\caption{Voltage profile in the transmission electrode layer for (a) $V_1 = 1$~V$_\textrm{RMS}$, $V_2 = 2$~V$_\textrm{RMS}$ ($1$~kHz) and (b) $V_1 = 2$~V$_\textrm{RMS}$, $V_2 = 1$~V$_\textrm{RMS}$ ($1$~kHz). (c) Zoom of the central area.}\label{Efield}  
\end{figure}

The total voltage distribution is numerically calculated by numerical methods considering a static approach of the LC (constant dielectric anisotropy, $\varepsilon_\textrm{o}$ and $\varepsilon_\textrm{e}$) and the whole structure. The results of Fig.~\ref{Efield}(a), corresponding to the voltage at the upper electrode plane for an AC signal of $1$~kHz with $V_1 = 2$~V$_\textrm{RMS}$ and $V_2=1$~V$_\textrm{RMS}$ demonstrate a voltage distribution that has a homogeneous transition in the active area. As seen in Fig.~\ref{Efield}(b), the use of inverted voltages in $V_1$ and $V_2$ demonstrates the possible use in generating both positive and negative topological charges. Finally, as can be seen in Fig.~\ref{Efield}(c), a central region remains uncovered with ITO due to the gap between slices, in this case, a circumference of $320$~{\textmu}m, which produces a voltage gap at the centre with the voltage dropping to zero. This effect is not noticeable in the experimental results and could be considerably reduced by improving the photolithography resolution. On the other hand, the effect of the gap between slices ($10$~{\textmu}m) also produces a slight voltage drop in this region, but considerably lower than the previous effect. The lines around the central gap observed in Fig.~\ref{Efield}(c) reveal a maximum drop of $5$~mV$_\textrm{RMS}$. Consequently, this effect should not be noticeable for LC with low birefringence and devices with low thickness (the phase step will not be high).

\begin{figure}[ht]
\centering\includegraphics[width=8cm]{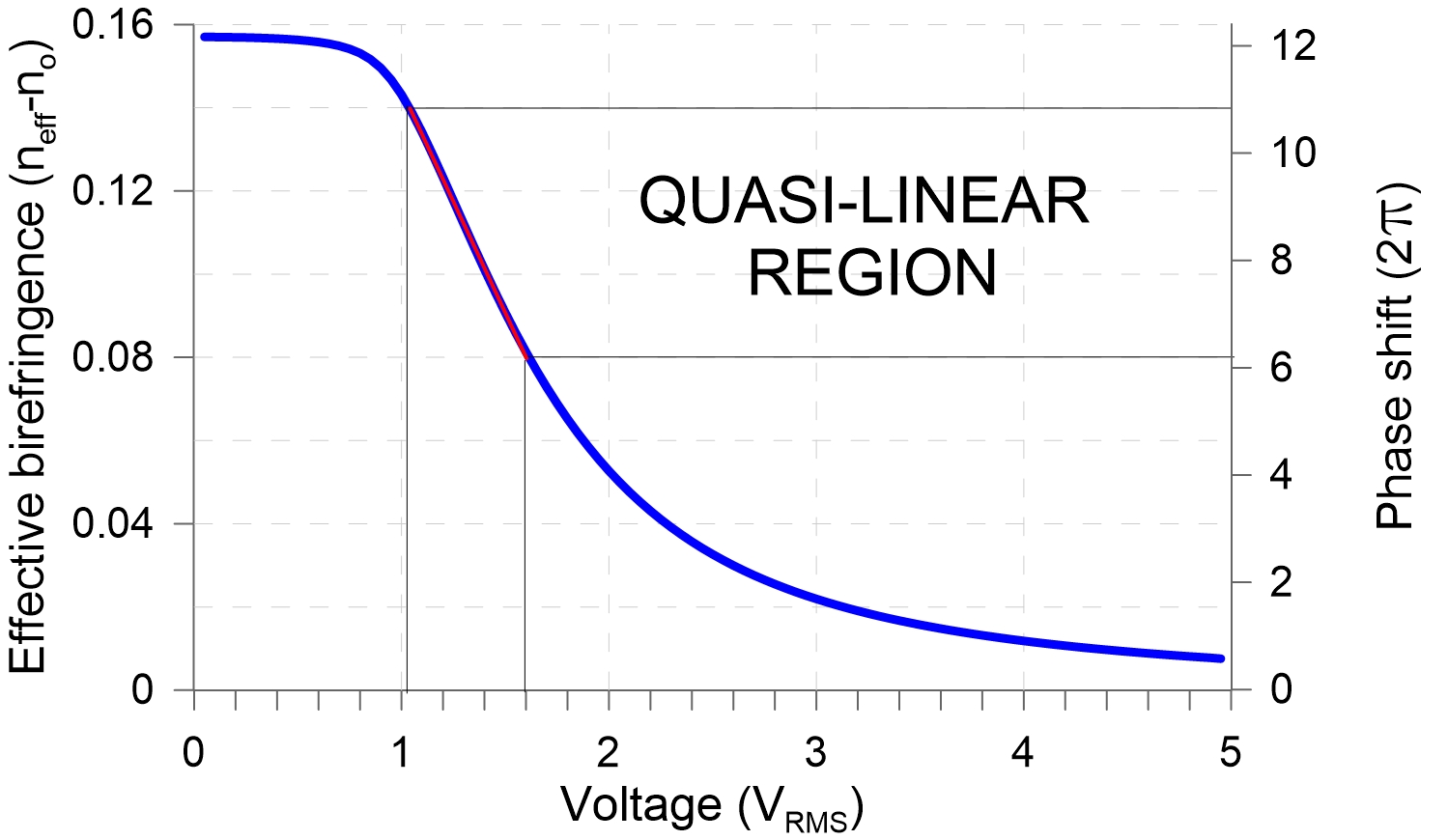}
\caption{Characteristic birefringence curve as a function of voltage for the nematic 6-CHBT.}\label{Birre}  
\end{figure}

The linearity of this electrical field distribution is key to maintaining a perfect vortex generator. However, the LC birefringence curve is non-linear, so this issue has to be carefully taken into account. A simulation of the LC birefringence response as a function of voltage (taking into account the 6-CHBT elastic constants and Frank-Oseen equations \cite{algorri2013}) is presented in Fig.~\ref{Birre}. The axis on the right represents the phase shift considering that the phase equals $\phi=\Delta n d/\lambda (2\pi)$~rad, the LC thickness is $d = 50$~{\textmu}m and $\lambda = 632.8$~nm. Traditionally, two characteristic regions in this curve have to be avoided. The first one is produced when the electrical energy overcomes the elastic free energy and the molecules start to move (just above the threshold voltage, $V_\textrm{th}$). The other one is produced when the molecules start to reach the perpendicular position (at the saturation voltage, $V_\textrm{sat}$). In this case, the range from $V_\textrm{th}$ to $V_\textrm{sat}$ is around $0.9$~V$_\textrm{RMS}$ to $3$~V$_\textrm{RMS}$, respectively. Close to these two values the birefringence still has a high non-linearity. Hence, a considerable margin has to be considered in order to operate in a quasi-linear region, in this case from $1$~V$_{RMS}$ to $1.6$~V$_{RMS}$. As a linear phase profile characterises an ideal SPP, this operating range has to be maintained. Considering this range (effective birefringence $\Delta n = 0.14-0.08 = 0.06$), the maximum linear phase shift is $4.74$~($2\pi$) rad, corresponding to a maximum theoretical topological charge of $l = 4$.

\section*{Setup and experimental results}

\begin{figure}[ht]
\centering
\includegraphics[width=12cm]{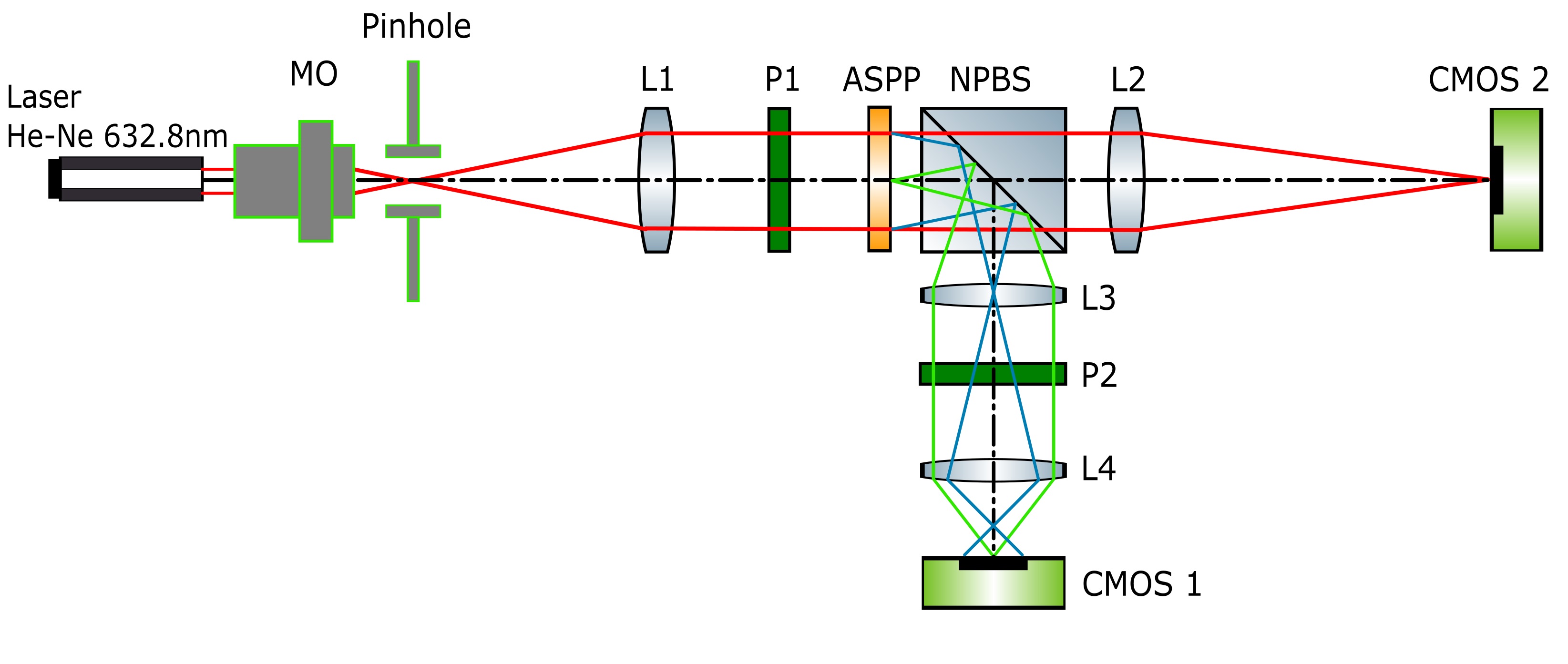}
\caption{Schematic of the optical system for measuring simultaneously the azimuthal change of phase retardation and the generated optical vortex beam of the ASPP.}\label{Setup}
\end{figure}

The optical system shown in Fig.~\ref{Setup} is used to simultaneously measure the azimuthal change of optical phase retardation characteristic of ASPP and the generated optical vortex beam. The light source for this system is an expanded and collimated laser beam with a wavelength of $632.8$~nm. A non-polarizing beam splitter (NPBS) is placed after the ASPP to divide the beam into two paths. One of these paths, reflected by the NPBS, is used to ensure the correct selection of the applied voltages ($V_1$ and $V_2$). 

This beam is directed onto the ASSP, which is positioned between two linear crossed polarisers (P1 and P2). These polarisers are oriented at $+45$\textdegree and $-45$\textdegree with respect to the LC optical axis, respectively. This setup allows for measuring the interference pattern between extraordinary and ordinary rays. The image of the ASPP is adapted onto the CMOS 1 camera (BASLER acA2500-60um) using two bi-convex lenses (L3, L4). From the data captured by this camera, the topological charge can be calculated by observing the bright-dark transitions (each transition corresponds to a $2\pi$ radian change). The other path, transmitted through the NPBS, is focused by the lens L2 onto the CMOS 2 camera (FLIR CM3-U3-13S2C-CS). This camera is located at the focal plane of the L2 lens. This configuration allows for capturing the optical vortex beam at the focal plane of the lens when the polariser P1 is parallel to the rubbing direction of the ASPP. When a light beam passes through a polariser whose polarisation axis is parallel to the LC optical axis, the light is affected by the effective refractive index of the molecules.

\begin{figure}[ht]
\centering
\includegraphics[width=15cm]{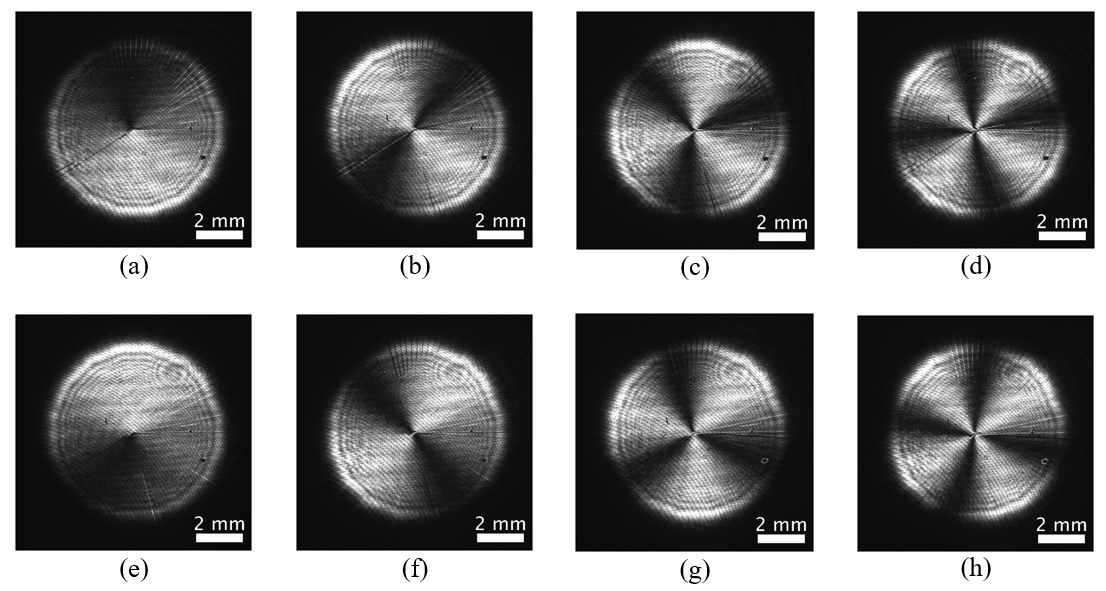}
\caption{Phase shift for different topological charges (a) $l = 1$, $V_1 = 1.6$~V$_\textrm{RMS}$, $V_2 = 1.46$~V$_\textrm{RMS}$,  (b) $l = 2$, $V_1 = 1.6$~V$_\textrm{RMS}$, $V_2 = 1.34$~V$_\textrm{RMS}$, (c) $l = 3$, $V_1 = 1.61$~V$_\textrm{RMS}$, $V_2 = 1.24$~V$_\textrm{RMS}$, (d) $l = 4$, $V_1 = 1.61$~V$_\textrm{RMS}$, $V_2 = 1.14$~V$_\textrm{RMS}$. (e)-(h) For $l = -1$ to $l = -4$ the inverted voltages in $V_1$ and $V_2$ are applied.}\label{Phase}
\end{figure} 

The results of the phase shift distribution can be observed in Fig.~\ref{Phase}. The LC layer has the ability to generate eight distinct OAM  modes, specifically from $l =\pm1$ to $\pm4$. As it was predicted in the birefringence study, minor alterations in the applied voltage can lead to substantial modifications in phase shifts. It is important to note that $2\pi$ multiples are necessary for the formation of optical vortices. Additionally, by minutely adjusting the applied voltage, the phase shift can be sustained at $2\pi$ (or a multiple thereof). In our case, to obtain perfect linear phase profiles one voltage is fixed around $1.6$~V$_\textrm{RMS}$ and the other shifts from $1.46$~V$_\textrm{RMS}$ (gradient of $0.14$~V$_\textrm{RMS}$ for $l=\pm1$) to $1.14$~V$_\textrm{RMS}$ (gradient of $0.47$~V$_\textrm{RMS}$ for $l=\pm4$).

\begin{figure}[ht]
\centering
\includegraphics[width=17.5cm]{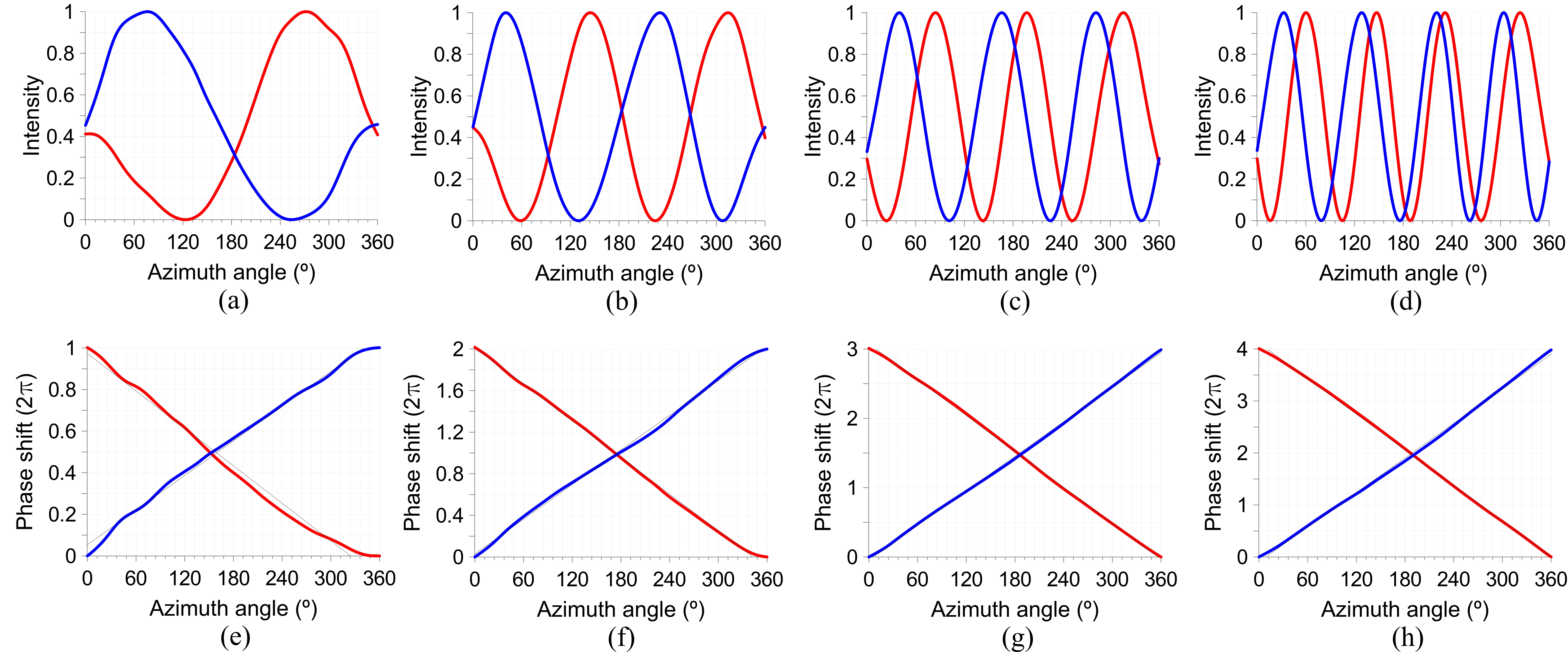}
\caption{(a)-(d) Intensity and (e)-(h) phase shift profiles for different topological charges. The applied voltages are (a),(e) $l = \pm 1$, $V_1 = 1.6$~V$_\textrm{RMS}$, $V_2 = 1.46$~V$_\textrm{RMS}$, inverted voltages in red, (b),(f) $l = \pm 2$, $V_1 = 1.6$~V$_\textrm{RMS}$, $V_2 = 1.14$~V$_\textrm{RMS}$, inverted voltages in red, (c),(g) $l = \pm 3$, $V_1 = 1.61$~V$_\textrm{RMS}$, $V_2 = 1.24$~V$_\textrm{RMS}$, inverted voltages in red, (d),(h) $l = \pm 4$, $V_1 = 1.61$~V$_\textrm{RMS}$, $V_2 = 1.24$~V$_\textrm{RMS}$ inverted voltages in red.}\label{Phaseprofile}
\end{figure}

A phase unwrapping procedure is carried out to demonstrate the linearity of the phase profile. The phase profile analysis was conducted through a series of steps. A median filter was initially applied to the raw images utilising a 15x15 pixel window. Subsequently, the weighted median angular intensity was computed at eight distinct radii, ranging from $100$ to $450$ pixels, with $1000$ points along the circle, Fig.~\ref{Phaseprofile}(a)-(d). The weight of the smaller circles was scaled down according to their radii. To ensure that the start and end points of the fitted curve matched the raw data, weights were added to the fit for the spline fitting of the intensity. This was crucial to ensure that we returned to the same intensity after traversing through all $360$\textdegree. Following this, peak-to-peak normalisation of the intensities was performed. This step was necessary as the phase between maximum and minimum intensity had to be $\pi$ for higher topological charges, and it facilitated the phase approximation.

As can be observed, for higher topological charges, there is a slight deviation from applying positive voltage gradient or negative voltage gradients (the absolute intensity signal is not perfectly inverted). This can be caused by some impurities at the LC, producing an undesired accumulation of ions and some hysteresis. Despite this, this effect does not affect the relative phase profiles, as shown below.

Lastly, the phase change between $0$ and $360$\textdegree~in that line was calculated following the Jones Matrix formalism. Specifically, it has to be taken into account that the vertical polarisation component is affected by the LC device and the radial linear phase deviates this component towards the beam axis. On the contrary, the horizontal linear polarisation component is unaffected and remains collimated. Thus, except for a constant phase term, the Jones matrix describing the ASPP can be written as Eq. \ref{eqJ}.
\begin{equation}\label{eqMASPP}
	M_\textrm{ASPP}(r) =\begin{bmatrix}
1 & 0\\
0 & \textrm{exp}[i\phi(r)]
\end{bmatrix},
\end{equation}
where $\phi$(r) is the radially dependent retardation which, ideally, should be linear with $r$ as $\phi(r) = k^\textrm{ASPP}_r r$ being $k^\textrm{ASPP}_r$ the radial wavenumber imparted by the ASPP device. Now, considering a second polariser (P2) oriented at 45\textdegree~behind the ASPP, the two polarisation components interfere and the Jones vector after the polariser reads
\begin{equation}\label{eqJ}
Jout(r) = \frac{1}{\sqrt{2}}\begin{bmatrix}
1 & 1\\
1 & 1
\end{bmatrix}\cdot M_\textrm{ASPP}(r)\cdot \frac{1}{\sqrt{2}}\begin{bmatrix}
1 & 1\\
\end{bmatrix}^T=\frac{1}{2}\{1+exp[\phi(r)]\}\begin{bmatrix}1\\1\end{bmatrix},
\end{equation}
where the first term corresponds to the Jones matrix of the second polariser and the third to the normalised Jones vector standing for the input beam ($T$ indicating the transpose vector). Considering this, the output field intensity is given by
\begin{equation}\label{eqInt}
I(r) = J^\dagger_\textrm{out}\cdot J_\textrm{out} = \cos^2[\phi(r)/2],
\end{equation}
where $\dagger$ indicates the Hermitian complex-conjugate vector.

To unwrap the phase, a MATLAB code was designed to transform the phase signal into a monotonically increasing sequence. This is achieved by initialising a new array of the same length as the phase signal, with the first element identical to the phase signal. A loop is then executed that traverses each element of the phase signal, starting from the second element. Within this loop, the absolute value of the derivative of the phase signal at the previous index is computed and added to the previous element of the new array. This process is repeated for all elements in the phase signal, resulting in a new signal that is monotonically increasing, regardless of the behaviour of the original signal. The results are shown in Fig.~\ref{Phaseprofile}(e)-(h). The phase profiles corresponding to negative topological charges are inverted and plotted in red. The obtained phase shifts are approximated to a linear fit (black line), demonstrating the low deviation from the ideal response.

The non-linearity issues, primarily observed in $l=1$ and $2$, are caused by the manufacturing process of the sample. Geometrical deformations in thickness cause an improper distribution, resulting in changes up to $\pi/2$ across the entire area. The maximum observed topological charge is $8$, but as $1$~V$_\textrm{RMS}$ is required, the birefringence curve is non-linear, producing undesired effects in the phase profile. Moreover, the inverted voltage is going to produce a different phase profile. To solve this issue and reach higher values of the topological charge a LC with a more linear birefringence curve will be required. Also, using a higher birefringence LC can solve this but a higher resolution lithography will be required to minimise the size of the central hole and the gaps between slices.  

\begin{figure}[ht]
\centering
\includegraphics[width=14cm]{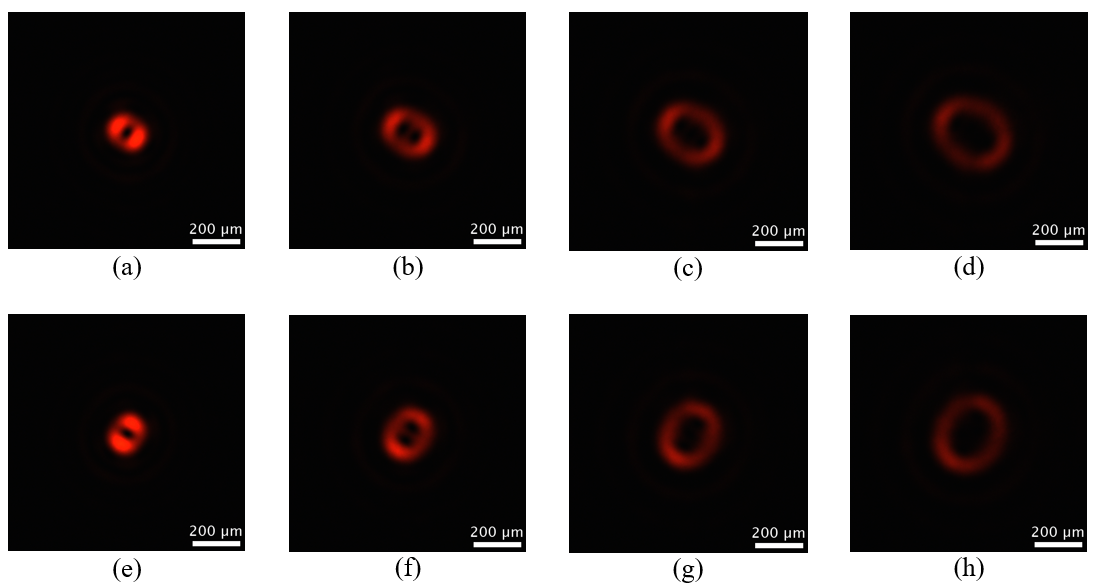}
\caption{Focal spot for different topological charges (a) $l = 1$, $V_1 = 1.6$~V$_\textrm{RMS}$, $V_2 = 1.46$~V$_\textrm{RMS}$,  (b) $l = 2$, $V_1 = 1.6$~V$_\textrm{RMS}$, $V_2 = 1.14$~V$_\textrm{RMS}$, (c) $l = 3$, $V_1 = 1.61$~V$_\textrm{RMS}$, $V_2 = 1.24$~V$_\textrm{RMS}$, (d) $l = 4$, $V_1 = 1.61$~V$_\textrm{RMS}$, $V_2 = 1.24$~V$_\textrm{RMS}$. (e)-(h) For $l = -1$ to $l = -4$ the inverted voltages in $V_1$ and $V_2$ are applied.}\label{PSF}
\end{figure}

Finally, insight into the topological charge of an optical vortex can be extracted from its point spread function (PSF), which is the impulse response of a focused optical system, meaning that it reveals how the light intensity is distributed at the focal spot. Fig.~\ref{PSF} presents the PSF at various voltages for a wavelength of $632.8$~nm. These results demonstrate that the topological charge can be controlled with precision. A slight variation of the voltage allows eight different modes to be achieved. The negative topological charge can be easily determined by interchanging the voltages at $V_1$ and $V_2$, as demonstrated before. 

\section*{Conclusions}
In summary, a novel optical vortex generator was proposed and experimentally demonstrated. Using this kind of ASPP, high-quality optical vortices with topological charges from $\pm1$ to $\pm4$ can be generated by only one ASPP. Owing to the continuous optical phase shift, this device is the best approximation to an ideal ASPP proposed to date. Moreover, the device is entirely reconfigurable (operating wavelengths and topological charges are tunable). The fabrication process is quite similar to that of a normal LCD cell, so it is low-cost and reliable. The device can be used in new applications (e.g., fibre optics communications or atom manipulation) to reduce the fabrication costs of existing devices and generate different OAM modes with improved light efficiency, simplicity, and the possibility of reconfiguration.

\section*{Author contributions statement}

J.F.A. conceived the idea and designed the device, J.F.A. and D.C.Z. conducted the simulations, P.M., M.F. and M.S. fabricated the device, T.J., P.M., A.P. and N.B. conceived the experiments, T.J., P.M. and N.B  conducted the experiments, J.F.A., T.J., P.M., A.P., D.C.Z. and N.B. analysed the results. J.F.A., T.J. P.M. and D.C.Z. wrote the manuscript, J.F.A., J.M.L.H., A.P. and N.B. supervised the work. All authors reviewed the manuscript.

\section*{Acknowledgements}

This work is part of the project PID2019-107270RB-C21 and PID2019-109072RB-C31 funded by MCIN/AEI/10.13039/501100011033 and FEDER ``A way to make Europe'', PDC2021-121172-C21 funded by MCIN/ AEI/10.13039/501100011033 and European Union ``Next generation EU''/PTR and project S2018/NMT-4326 funded by the Comunidad de Madrid and FEDER Program. N. Bennis and A. Spadlo also acknowledge research project UGB 22-791 (Military University of Technology) and NAWA PROM projekt nr POWR.03.03.00-00-PN13/18 under European Social Fund. J.F.A. received funding from Ministerio de Ciencia, Innovaci{\'o}n y Universidades of Spain under Juan de la Cierva-Incorporaci{\'o}n grant. 

\section*{Additional information}

The authors declare no conflicts of interest.

\section*{Availability of data and materials}

The datasets used and/or analysed during the current study are available from the corresponding author upon reasonable request.


\end{document}